\documentclass[aps,twocolumn,prl,groupedaddress,showpacs]{revtex4-1}
\usepackage[T1]{fontenc}
\usepackage[latin9]{inputenc}
\usepackage{times}
\usepackage{color}
\usepackage{xspace}
\usepackage{amssymb,amsmath}
\usepackage{amsbsy}
\usepackage{graphicx}
\usepackage{multirow}
\usepackage{rotating}
\usepackage{microtype}
\usepackage{tabularx}
\usepackage{epstopdf}
\usepackage{float}
\usepackage{mathrsfs}
\usepackage{hyperref}
\usepackage[normalem]{ulem}
\def\be{\begin{equation}}
\def\ee{\end{equation}}
\def\e#1{\label{#1}\end{equation}}
\def\bea{\begin{eqnarray}}
\def\eea{\end{eqnarray}}

\def\ket#1{{|#1\rangle}}
\def\bra#1{{\langle#1|}}

\def\braket#1{\langle{#1}\rangle}
\begin{document}
\title{Heralded Control of Mechanical motion by Single Spins}
\author{D. D. Bhaktavatsala Rao }
\author{ S. Ali Momenzadeh}
\author{J\"{o}rg Wrachtrup}
%\email{wrachtrup@physik.uni-stuttgart.de}

\affiliation{3. Physikalisches Institut, Research Center SCOPE, and MPI for Solid State Research, University of Stuttgart, Pfaffenwaldring 57, 70569 Stuttgart, Germany
}%
\date{\today}
\begin{abstract}
We propose a method to achieve high degree control of nanomechanical oscillators by coupling their mechanical motion to single spins. By manipulating the spin alone and measuring its quantum state heralds the cooling or squeezing of the oscillator even for weak spin-oscillator couplings. We analytically show that the asymptotic behavior of the oscillator is determined by a spin-induced thermal filter function whose overlap with the initial thermal distribution of the oscillator determines its cooling, heating or squeezing.  Counterintuitively, the rate of cooling dependence on the instantaneous thermal occupancy of the oscillator renders robust cooling or squeezing even for high initial temperatures and damping rates.  We further estimate how the proposed scheme can be used to control the motion of a thin diamond cantilever by coupling it to its defect centers at low temperature.
 \end{abstract}
%\pacs{tttt}
\maketitle

Over the past decade ground state cooling of mechanical oscillators have paralleled the laser cooling methods for atoms, thus allowing to realize quantum mechanical effects even at microscopic length scales \cite{ref1}. With this ability these devices can be used in a variety of applications ranging from quantum metrology to quantum information processing \cite{hybrid}. Cooling (micro) nanomechanical oscillators (NMO) by laser induced radiation pressure force has received huge attention both theoretically and experimentally which lead to the observation of ground state cooling of mechanical motion \cite{ref2}, quantum back action \cite{ref3} and other nonclassical effects like squeezing \cite{ref4}.  Recent advances in solid-state materials hosting single localized spins (emitters) are also shown to be promising candidates both for controlling the dynamics of the oscillator and also induce interaction (entanglement) between distant spin by the dynamical back-action of the NMO\cite{ref5}.

Solid state spins in diamond are robust in terms of their good spin coherence properties, high degree spin control \cite{ref6} and a well-resolved optical spectrum at low temperatures \cite{ref7} that allows optical excitation to various levels that can either initialize or readout the electronic spin state with fidelities exceeding $98\%$ f\cite{ref8}. On the otherhand, they suffer from weak coupling to external spins, incidents photons \cite{ref9} and phonons \cite{exp2}. Even with such weak couplings, projective readout techniques, can be used achieve deterministic nuclear spin state preparation \cite{ref12}, entanglement between solid state qubits mediated by photons \cite{ref13}, transfer and storage of single photon states in nuclear spins \cite{ref14}. In this work we employ these post-selection techniques to achieve robust control of mechanical modes of a NMO that are weakly coupled to the host spins.

Coupling single spins to the mechanical motion of a NMO has already been demonstrated \cite{exp1, exp2, exp3, exp4}. Due to the large thermal occupancy of these modes even at low temperatures, realizing gates between distant spins coupled to a common mechanical mode is a nontrivial task \cite{ref5}. Hence one needs to cool down or squeeze these modes to a high degree to observe any coherent effects or perform gates between distant spins that are coupled to a common mode of NMO. However, with the extremely small spin-phonon coupling achieving such tasks could be challenging. In other words, to achieve coherent coupling between a spin and a phonon  the spin cooperativity $C_S =g^2T_2/\gamma$, should be be greater than unity \cite{lukin}. This can be possible only if the inverse spin coherence time, $1/T_2$, and the dissipation rate of the oscillator $\gamma$ are smaller than the spin-phonon coupling $g$. In this work we will show that even when $g \le \gamma, T_2$, repetitive projections of the spin to a specific state allows a filtering of unwanted thermal occupancies of the oscillator, thereby allowing us to cool down, heat or squeeze its motional degrees of freedom.

The proposed scheme is based on the conditional evolution of the NMO by repetitively post-selecting the spin state dynamics to which it is coupled. When coupling quantum systems of different Hilbert space dimensions, i.e., a two-level spin with an $N$-level oscillator, a short time evolution that looks like a perturbation on the spin state observables can lead to a dramatic change on the oscillator dynamics. Though, repeatedly finding the spin in a given state appears like an effective freezing of its dynamics (Zeno effect), on the contrary can drive the other system quite far from its equilibrium state \cite{naka, klaus}. In this work we will exactly solve the dynamics of the NMO conditioned on the repetitive measurement of a solid state spin to which it is coupled, and extract a measurement induced nonlinear cooling rate that leads to a rapid near ground state cooling of the oscillator modes.

%Coupling and controlling weak spin-spin interactions has been a well developed technique in magnetic resonance studies. With an appropriate stroboscopic control of the spin one can design a noise filter function such that coupling can be acheived only at a given frequency allowing one to observe coherent effects even for very weak interactions. Similar to these ideas we will show how one could design a spin-controlled (thermal) filter function for the mechanical oscillator that filters out unwanted thermal occupancies allowing it to cool down, heat up or squeeze its motional degress of freedom.

%To overcome the exterme weak coupling between different quantum entities and yet observe nonclassical effects, recently heralded (post-selection) techniques are being widely used. For example to observe spin photon entnaglement, (or entanglement between spins mediated by photons) where the coupling between the spin and single photon is $\sim 10^{-4}Hz$, one rather relies on counting the succesful events of entanglement over many experimental runs, rather reading the fidelities in a single experiment. With an appropriate heralding this procedue could also be made quasi-determinsitic as a signal is received (though with a low probability) whenever the experiment is succesful. . In this work we will harness the good spin properties of NVC to achieve robust control of weakly couled mechanical modes.

We shall consider the geometry shown in Fig.1 (a), i.e., a one-sided clamped microcantilever with single spins implanted in it. In the presence of a large magnetic field gradient, these spins (two-level systems) couple to a position-dependent magnetic field i.e., the spins experience a phase shift of their energy states that depends on the position of the cantilever. The coupling between single spin and a given mechanical mode of frequency $\omega_m$ is determined by the zero point motion of the oscillator and the magnetic field gradient \cite{Rabl}. The Hamiltonian describing the dispersive interaction between the spin and the NMO (in units of $\hbar = 1$) is given by
\be
H=\sum_m [\omega_m a^\dagger_m a_m + g_mS^z (a_m +  a^\dagger_m)] + \Omega(t)S^x,
\ee
where $g_m$ is the coupling between the spin ($S$) and the $m^{\rm{th}}-$mode of the oscillator ($a_m$) that has a frequency $\omega_m$. The stroboscopic control of the spin is determined by the time-dependent function $\Omega(t)$. In addition to this unitary coupling the spin and the oscillator modes suffer nonunitary decay processes which we will consider in the later part of this paper.
Under evolution governed by the above Hamiltonian with a stroboscopic spin control i.e., $\Omega(t) =\pi \sum_k \delta(t-\tau_k)$, the time-evolution operator takes the simple form
\be
U(t) = \mathcal{D}_+(t)\ket{1}\bra{1}+\mathcal{D}_-(t)\ket{0}\bra{0},
\ee
where $\ket{1(0)}$ are the eigenstates of the spin operator $S^z$ and the multimode displacement operators $\mathcal{D}_\pm = \exp({\pm i\sum_m g_m(F_t (\omega_m)a_m + F^*_t (\omega_m)a^\dagger_m)/\omega_m})$ \cite{uhrig}.
The (mode) filter function, $F_m(t)$, is  induced by the applied control which for equally spaced $\pi$-pulses intervals (i.e.,$\tau_k=\tau \equiv \pi/\omega$) takes the well-known form, $F_\tau(n_c,\omega) =4\tan^2[\omega t/(2n_c+2)]\cos^2(\omega t/2)$.
In the limit of large number of $\pi$-pulses, $n_c$, the filter function $F_m(t) \approx 2n_c\delta(\omega-\pi/\tau)$, i.e., coupling to single mode (or to modes with a frequency that are integral multiples of $\omega$ are only relevant). The total time is now expressed in units of $\tau$ and the dynamics will be determined by a single dimensionless parameter $\lambda = 2g n_c/\omega$. 

The pulse sequence for implementing the proposed scheme is shown in Fig. 1 (c). Starting from the initial spin state $\ket{0}$, we prepare a superposition state using the first $\pi/2$ pulse and allow it to evolve for a $t = n_c\tau$ with stroboscopic interruptions at intervals $\tau$. The spin is then projected back to the energy basis ($\ket{1(0)}$) with the final $\pi/2$ pulse. We now readout the state of the spin optically and note the measurement result. Owing to a successful measurement result i.e., finding the spin in the state $\ket{0}$, the procedure is repeated $M$ times, as shown in Fig.1. Since the time $t$ is arbitrary the probability of obtaining a successful measurement is always below unity. Given this conditional spin state dynamics the evolution of the NMO that is initially in thermal equilibrium is greatly effected which we analyze below.

\begin{figure}
\begin{center}
\includegraphics[width=0.5\textwidth]{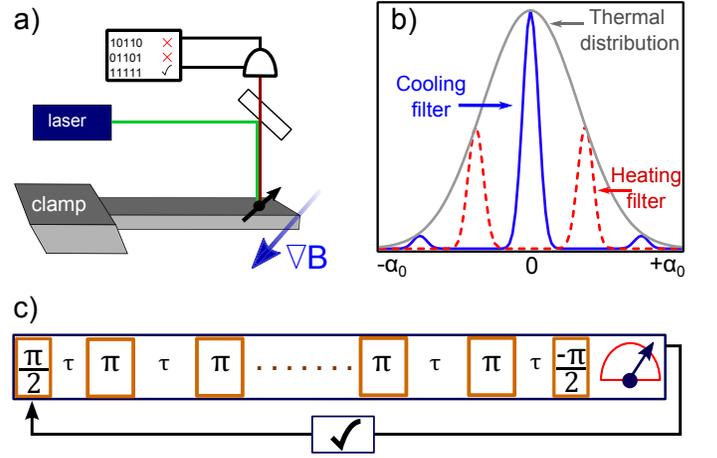}
\end{center}
\label{level}
\vspace{-5mm}
\caption{(a) Schematic illustration of the control procedure detailed in the paper. A single-side clamped NMO hosting single spins (black) experiencing a gradient magnetic field $B$ (blue) is shown. Further the heralding mechanism is depicted by using a laser for optical readout of the solid-state spin where the emitted light is collected on a photo-detector. Also shown is the readout sequence with '$1$' signifying the successful readout (projection) of the spin state and '$0$' the unsuccessful event. (b) Schematic illustration of the initial thermal distribution (grey solid-line) of the oscillator and the spin-controlled filter function that lead to either cooling (blue-solid line) or heating of the oscillator (red-dashed line).
(c) The total pulse sequence for the heralded control of the oscillator is shown. }
\end{figure}
For an oscillator in its thermal state $\rho_B =  \frac{1}{\sum_M e^{-\beta n}}\sum_M e^{-\beta n}\ket{n}\bra{n}$, with an initial thermal occupancy $n_\omega = Tr[\rho_B a^\dagger a]$, the effect of the spin-phonon coupling would lead to random displacements governed by the operators $\mathcal{D}$ as shown in Eq. (2). As we are dealing with the displacement operators, it is instructive to work in the coherent basis where the above-mentioned thermal state can be rewritten as 
\be
\rho_B(0) = \int \mathcal{P}_0(\alpha) \ket{\alpha}\bra{\alpha}d^2 \alpha, ~\mathcal{P}_0(\alpha) = \frac{1}{\pi n_\omega}e^{-|\alpha|^2/n_\omega}.
\ee
Upon obtaining a successful spin state measurement, the oscillator is projected onto the state 
\be
\rho_B(t) = \frac{V\rho_B(0)V^\dagger}{{\rm Tr}[V\rho_B(0)V^\dagger]}.
\ee
where $V(t)= \frac{1}{2}(\mathcal{D}_+(t) +\mathcal{D}_-(t))$. The success probability for this projection is then simply given by ${{\rm Tr}[V\rho_B(0)V^\dagger]}$. Now $\rho_B(t)$ would be the initial state of the oscillator for the next repetition and evolving again for a time $t$, a successful measurement of the spin will project the oscillator onto the state $\rho_B(2t) = \frac{V\rho_B(t)V^\dagger}{{\rm Tr}[V\rho_B(t)V^\dagger]}$.
After $M$-successful projections of the spin to its desired state the oscillator can again be expressed in the coherent basis as given in Eq. (3), where the $\mathcal{P}$-function is now modified as
\begin{eqnarray}
\mathcal{P}_M(\alpha) = {\mathcal{P}_0(\alpha) }G^\alpha_M(\epsilon,\lambda),
\end{eqnarray}
where 
\be
G^\alpha_M(\epsilon,\lambda) =C_M \prod_{k=1}^N{\rm Re}\left[\exp\left(\frac{\lambda (e^{i\epsilon t}-1)}{\epsilon}\left\lbrace\alpha e^{i(k-1)\epsilon t}+\alpha^*e^{-ik\epsilon t}\right\rbrace\right)\right]. 
\ee
acts like the spin-induced thermal filter function (see Suppl. Info) by suppressing unwanted excitations in the phase space of the oscillator.
In the above equation $\epsilon$ determines the slight off-resonant driving of the spin i.e., the evolution of the spin is interrupted stroboscopically at intervals of $\tau = \pi/(\omega - \epsilon)$, and $C_M$ is determined from the normalization condition of the oscillator state i.e., $Tr[\rho_B(Mt)] =1$. In Fig. 2 (a) - (b) we plot the $\mathcal{P}$-function in the complex space of $\alpha$ which show the initial thermal distribution, the squeezed distribution for resonant driving and a narrowed thermal distribution for off-resonant driving of the spin.  These confirm the above described picture of the role of a spin-induced filtering in the oscillator's Hilbert space to achieve a given task. We also confirm the above analysis by performing exact numerical diagonalization of the dynamics generated by the Hamiltonian given in Eq. (1). These results are shown in Fig. 2 (d) - (f).

The key aspect of this measurement control is to obtain a high probability for finding the spin in a given state ($\ket{0}$). To obtain such high probability, the effective spin-oscillator coupling should satisfy the condition $\lambda^2n_\omega < 1$. With this condition it is guaranteed that probability of finding the spin close to its initial state is much higher than $50\%$ during the first evolution cycle i.e, over a time $t$. This in turn also indicates that for higher temperatures it is better to have smaller effective couplings $\lambda$ (which is possible by adjusting the number of control pulses $n_c$). After the first successful projection of the spin, if the oscillator has also been projected onto a state with smaller thermal occupancy then the above condition guarantees that the success probability for the next heralding event will be much higher than the previous one. As the oscillator gets gradually cooled each heralding event increases the probability for the next one, thus resulting in a cascading behavior as shown in Fig. 2d where an exponential convergence to unity will be observed with either cooling or squeezing the oscillator. As the cooling procedure is not deterministic one can only find the event rate of achieving a ground state cooled oscillator state. For example, the net success probability (event rate) to achieve the cooling shown in Fig. 2 (f) is $0.125$ i.e., if a single experiment is described by $M$ repetitive spin-state measurements (see Fig. 1) then for every $8$ experiments one can find success in all $M$ measurements, thereby, driving the oscillator towards maximally cooled/squeezed state.

\begin{figure}
\vspace{-5mm}
\begin{center}
\hspace{-5mm}
\includegraphics[width=1.\columnwidth]{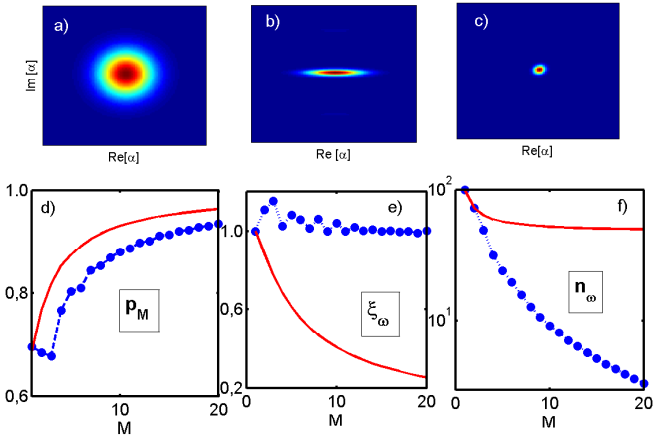}
\end{center}
\label{level}
\vspace{-5mm}
\caption{The $\mathcal{P}_M$-function is plotted in the complex space of $\alpha$ for (a) initial thermal state, (b) squeezed state obtained under resonant driving of the spin ($\epsilon = 0$) and (c) thermally cooled state for off-resonant driving of the spin $\epsilon=0.1\lambda\omega$. We have used $\lambda = 0.25$, and $M=10$. (d). The probability of successful heralding is plotted as function of number of heralding events $M$. (e) The squeezing of the oscillator position and (f) the thermal occupancy of the oscillator is plotted as a function of $M$. In plots (d)-(f) we compare the cases with different periodicity for the applied $\pi$-pulses: resonant driving $\tau = \pi/\omega$ (red-solid line) and slight off-resonant driving $\tau = \pi/(\omega + \epsilon)$ (blue circles).
For the above plots the parameters used in the simulation are $g/\omega = 2.5 \times 10^{-4}$, the M-dependent control pulses $n_c (M) = 10^2M^{0.25}$ and the offset $\epsilon = 20g$. The exact numerical diagonalization is performed for an oscillator with $10^3$ basis states.  }
\end{figure}

For cooling down the oscillator all the way to its ground state we have shown that the spin should be driven at a frequency that is slightly off-resonant by $\epsilon \sim g$ from the mode frequency $\omega$. The physical reason for this comes from the preferential basis chosen by the conditional dynamics on the phase space of the oscillator. For resonant driving the displacement operator, $V(t) = \cos [\lambda (a+a^\dagger)]$, is diagonal in the position basis of the oscillator and hence repetitive projections of the oscillator onto this basis will stabilize the system in one of the eigenstates of the position operator, thereby squeezing it maximally (see Fig. 2 (b), 2 (e)). On the other hand, with slight off-resonant driving, the displacement operator $V(t) \approx {\rm Re}(\exp [\lambda (c_0 a+ c_0^*a^\dagger)])$ (where $c_0$ is some complex number) has both position and momentum components, and this does not allow the oscillator to stabilize in either of the basis. Therefore the system continues to remain diagonal in the energy basis with thermal occupancy decreasing for increasing $M$ (see Fig. 2 (c), 2 (f)).

As the initial decay of the thermal occupancy is similar for both resonant and off-resonant driving (see Fig. 2 (f)) we will obtain the ground state cooling rate by analyzing the resonant case as it has a comparatively simpler form for the operator $V(t) =\cos [\lambda (a+a^\dagger)]$. Using this, and for $\lambda^2 n_\omega < 1$, one can evaluate $n_\omega(t) \equiv {\rm Tr}[V(t)a^\dagger a V(t) \rho_B(0)] =n_\omega(0)[1-\lambda^2(2n_\omega(0)+1)]$. The cooling behavior for $M$ measurements can then be approximated by (see Suppl. Info)
\be 
n_\omega((M+1)t) \approx {n_\omega(Mt)}e^{-2\lambda^2n_\omega(Mt)}.
\ee
From the above equation, one can the effective cooling rate $\gamma_M \approx \left(\frac{4g^2}{\omega}\right) \braket{a^\dagger a}_M$.
Surprisingly, the cooling rate itself depends on the instantaneous thermal occupancy of the oscillator i.e., higher the initial thermal occupancy higher is the decay (cooling rate). This cooling rate decreases with increasing $M$, eventually reaching to the single phonon limit, which is $\frac{4g^2}{\omega}$. We further find that the mode occupancy can be reduced on an average by approximately $70\%$ in each repetition cycle, hence the  quantum speed limit of cooling the oscillator to its ground state is possible by $M \approx 2\log_2 (n_\omega)$ successful spin projections.  The resolved side-band cooling methods generally employed for cooling the NMO \cite{Rabl} will only reduce the oscillators thermal occupancy in unit steps i.e., $\ket{0}\ket{n}\rightarrow \ket{1}\ket{n-1}$ over a time-scale of $\sim 1/g$, indicating that more optical cooling cycles for the spin have to be employed to reduce the oscillator thermal occupancy to half its initial value and $g > \Gamma$ to ensure cooling. On the contrary, the current method importantly shows that one successful projection of the spin can reduce the thermal occupancy drastically and its success probability is bounded by $0.5$. 
%For completeness, we have also verified the analytical cooling rate given in Eq. (7) with that obtained from exact numerical diagonalization, and they show quite a good agreement (see Suppl. info). 
\begin{figure}
\vspace{-5mm}
\begin{center}
\includegraphics[width=0.5\textwidth]{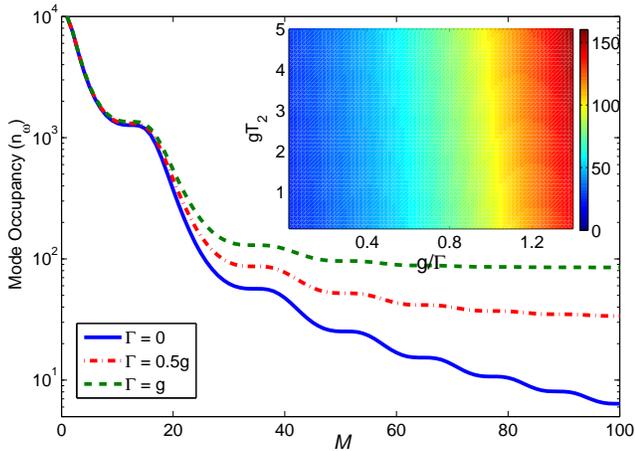}
\end{center}
\label{level}
\vspace{-5mm}
\caption{We show the cooling of the oscillator as a function of heralding round $M$ for various damping rates of the oscillator. In the inset we show the final steady state population as a function of $\Gamma$ and the spin coherence time $T_2$ for a fixed $M = 50$.}
\end{figure}

As the spin is measured frequently at time intervals of $t$, the spin dephasing/decay processes are only relevant for time $t$. At low ($\sim 4K$) temperatures and with periodic $\pi$-pulses used in our scheme (Fig. 1c), the non-unitary spin processes become less important. On the other hand, the dissipation of the oscillator plays a prominent role in the cooling procedure. To include the damping effects of the oscillator, $\Gamma \ll \omega$, into the analysis all we need is to replace $\epsilon$ in Eq. (5) with $\epsilon + i\Gamma$ (see Suppl. Info). Under this assumption there are two competitive effects: (i) the spin-induced cooling and (ii) thermalization of the oscillator. These two processes allow the system to reach a steady state where the final thermal occupancy of the oscillator depends on the ratio between the spin-oscillator coupling $g$ and the dissipation rate $\Gamma$. We show this behavior in Fig. 3. For finite $\Gamma$, the thermal occupancy of the NMO stabilizes with increasing $M$, reaching a quasi-steady state where the cooling and heating rates are balanced. In the inset we show the dependence of this steady state value both as a function of spin coherence time $T_2$ and $\Gamma$. As described earlier, $T_2$ has almost no net effect on the final thermal occupancy, it only influences the probability of the spin state measurements. For $T_2 \rightarrow 0$, the success probability in each round becomes $0.5$ which means that in a total of $M$ repetitions the probability to observe a ground state cooling event is  $\le 1/2^M$. Hence, the steady state value is only determined by the parameters $\lambda$ and $\Gamma$, and only has a weak dependence on $T_2$ as shown in the inset of Fig. 3. Comparing the cooling rate $\gamma_M$ with the damping rate $\Gamma$ one can see that the total number of measurements $M$ that contribute to cooling is determined by $\gamma_M/\Gamma > 1$.

To experimentally implement the above described heralded control of mechanical motion, we will consider a single-side clamped diamond cantilever of dimensions $(l, w, t) = (20, 1, 0.2)\mu$m, with a fundamental frequency of $\sim 10$ MHz \cite{book}, and the zero point fluctuation of $10^{-14}m$. A magnetic (AFM) tip held at a distance of $10$nm from the surface of the diamond producing a field gradient of $2~$G/nm, will result in the spin-phonon coupling of $g \sim100$Hz \cite{Rabl}. A Nitrogen Vacancy center in diamond has a spin one (three-level) ground state of which the two energetically low spin states form the two-level spin $S$. These centers are found at a depth $> 10$nm and can have long $T^*_2 \sim \mu$s. With dynamical decoupling sequences for example the one shown in Fig. 1(c), the spin coherence time can be extended to $T_2 > 10$ms at cryogenic temperatures $(4K)$. At such low temperatures and in high vacuum environment where the experiments are performed the cantilever can have quality factors $\sim10^5$ \cite{exp3}. In addition to this finite damping rate of the fundamental mode of the NMO, the readout laser and the applied microwave control can also heat up the diamond, and hence contribute to an additional heating rate $\Gamma_h$. With optimal design of microwave resonators of high $Q$,  and using resonant excitation at $637nm$ with a very low laser powers of $~100$nW  for spin state readout, $\Gamma_h$ could also be drastically reduced.  One can also employ the cooling procedure for other geometries suggested in \cite{Rabl}, where a cantilever carrying a magnetic tip is cooled by the spins embedded in a large diamond membrane close to the tip to avoid additional heating rate $\Gamma_h$. 

In conclusion we show here that by post-selecting the spin dynamics the evolution of the NMO coupled to it could be steered far from its equilibrium either towards a cooled or a squeezed state. For ultra weak couplings i.e., $g < T_2, \Gamma$, where the spin cooperativity is close to zero the heralded spin control method proposed here offers an alternative for cooling the NMO via the single spin. The successive spin state measurements heralds the cooling/squeezing of the NMO. Though probabilistic, the event rate to achieve near ground state cooling is quite high due to the dependence of the cooling rate on the instantaneous thermal occupancy. These methods can be further used to achieve heralded entanglement between distant spins (or perform gates) that are weakly coupled to a common mode of NMO's.

\begin{acknowledgements}
        We would like to acknowledge the financial support by the ERC project SQUTEC, DFG (FOR1693), DFG SFB/TR21, EU DIADEMS, SIQS, Max Planck Society and the Volkswagenstiftung.
\end{acknowledgements}

%D. W. C. Brooks, et al., Nature 488, 476 (2012).

\end{document}